\documentclass[twocolumn,showpacs,aps,prl,letterpaper,superscriptaddress]{revtex4}

\usepackage{graphicx}
\usepackage{dcolumn}
\usepackage{amsmath}
\usepackage{epsfig}
\usepackage{subfigure}

\usepackage{ifthen}

\input babarsym
\newcommand{\BABARPubYear}    {05}
\newcommand{\BABARPubNumber}  {034}

\newcommand{\SLACPubNumber} {11358}
\newcommand{\LANLNumber} {0507069}

\newcommand{\btaunu}     {\ensuremath{\Bp \to \taup \nut}\xspace}
\newcommand{\BRbtaunu}   {\ensuremath{\BR(\Bp \to \taup \nut)}\xspace}

\newcommand{\eextra} {\ensuremath{E_{\mathrm{extra}}}}

\newcommand{\eLabGam} {\ensuremath{E_{\gamma}}}
 
\newcommand{\btodszlnu} {\ensuremath {\B^- \to D^{*0} \ell^- \nulb}}
\newcommand{\dszlnu} {\ensuremath {D^{*0} \ell^- \nulb}}

\newcommand{\enunu} {\ensuremath {e^+ \nu_e \nutb}}

\newcommand{\mununu} {\ensuremath {\mu^+ \nu_{\mu} \nutb}}

\newcommand{\pinu} {\ensuremath {\pi^+ \nutb}}

\newcommand{\pipiznu} {\ensuremath {\pi^+ \pi^{0} \nutb}}

\newcommand{\threepinu} {\ensuremath {\pi^+ \pi^{-} \pi^{+} \nutb}}

\newcommand{\bcount} {\ensuremath {(231.8 \pm 2.6) \times 10^{6}} }

\newcommand{\onlumi} {\ensuremath {210.6 \invfb}}

\newcommand{\offlumi} {\ensuremath {21.6 \invfb}}

\newcommand{\eeqq} {\mbox{$e^{+}e^{-} \to q \overline {q}$}}
\newcommand{\Ebeam}       {\mbox{$E_{\rm beam}$}}
\newcommand{\cosBDsl}       {\mbox{$\cos \theta_{B,D^{*0}\ell}$}}

\newcommand{\eetautau} {\mbox{$e^{+}e^{-} \to \tau^{+} \tau^{-}$}}
\newcommand{\tagBlep}        {\mbox{$B^-_{\rm sl}$}}

\newcommand{\etal} {{\it et al.}}

\def\figurebox#1#2#3{%
    \def\arg{#3}%
    \ifx\arg\empty
    {\hfill\vbox{\hsize#2\hrule\hbox to #2{\vrule\hfill\vbox to #1{\hsize#2\vfill}\vrule}\hrule}\hfill}%
    \else
    {\hfill\epsfbox{#3}\hfill}%
    \fi}

\begin{document}

\preprint{\babar-PUB-\BABARPubYear/\BABARPubNumber}
\preprint{SLAC-PUB-\SLACPubNumber}

\begin{flushleft}
\babar-PUB-\BABARPubYear/\BABARPubNumber\\
SLAC-PUB-\SLACPubNumber\\
hep-ex/\LANLNumber\\
\ \\
\end{flushleft}

\title{
{\large \bf  
A Search for the Decay \boldmath{\btaunu}} 
}

%
\author{B.~Aubert}
\author{R.~Barate}
\author{D.~Boutigny}
\author{F.~Couderc}
\author{Y.~Karyotakis}
\author{J.~P.~Lees}
\author{V.~Poireau}
\author{V.~Tisserand}
\author{A.~Zghiche}
\affiliation{Laboratoire de Physique des Particules, F-74941 Annecy-le-Vieux, France }
\author{E.~Grauges}
\affiliation{IFAE, Universitat Autonoma de Barcelona, E-08193 Bellaterra, Barcelona, Spain }
% ...
\collaboration{The \babar\ Collaboration}
\noaffiliation

\date{\today} 

\begin{abstract}
We search for the rare leptonic decay $\btaunu$ in a sample of 
$232 \times 10^6$ $\BB$ pairs collected with the \babar\ detector at the SLAC
PEP-II \B-Factory. 
Signal events are selected by examining the properties of the $B$ meson
recoiling against the semileptonic decay \btodszlnu.  
We find no evidence for a signal and set an upper limit on the branching 
fraction of $\BRbtaunu < 2.8 \times 10^{-4}$ at the 90\% confidence level.
We combine this result
with a previous, statistically
independent \babar\ search for \btaunu\ to give an upper limit of
$\mathcal{B}(\btaunu) < 2.6 \times 10^{-4}$ at the 90\% confidence level.
\end{abstract}

\pacs{13.20.He, 14.40.Nd, 14.60.Fg} 

\maketitle



In the Standard Model (SM) the purely leptonic decay $\btaunu$~\cite{bib:cc}
proceeds via the annihilation
of the \bbar\ and \u\ quarks into a virtual \W\ boson. 
Its amplitude is proportional to the product of the 
Cabibbo-Kobayashi-Maskawa 
(CKM) matrix~\cite{bib:CKM} element \Vub\ and the \B\ meson decay 
constant \fsubb.
The SM branching fraction is given by~\cite{bib:BaBarPhysBook}:
\begin{equation}
  \BRbtaunu  =  \frac{G^2_F m_B}{8 \pi} m_{\tau}^2 
        \left( 1 - \frac{m^2_{\tau}}{m^2_{B}} \right)^2 f^2_{B} \Vub
        ^2 \tau_{B},
  \label{eq:brsm}
\end{equation}
where $G_F$ is the Fermi coupling constant, $m_{\tau}$ and $m_{B}$ 
are the $\tau^{+}$ lepton and \Bu\ meson masses, 
and $\tau_{B}$ is the \Bu\ lifetime. 
The branching fractions for
$\Bu \to \ep \nue$ and $\Bu \to \mup \num$ are helicity-suppressed by
$m_{\ell}^2 / m_{B}^2$, where $m_{\ell}$ is the mass of $\ep$ or $\mup$.
Using the value of 
$\Vub = (3.67 \pm 0.47)\times 10^{-3}$~\cite{bib:pdg2004} and the
lattice QCD calculation of 
$\fsubb = (0.196 \pm 0.032) \gev$~\cite{bib:Kronfeld}, we
determine an expected value of 
$\BRbtaunu = (9.3 \pm 3.9) \times 10^{-5}$.
Currently, our best knowledge 
of \fsubb comes from theoretical
calculations, with a current theoretical uncertainty of 
roughly 16\%~\cite{bib:Kronfeld}. 
Observation of $\btaunu$ could provide the first direct measurement of \fsubb.
The ratio of $\BRbtaunu$ and $\Delta m_d$, 
the difference in heavy and light
neutral $B_d$ masses~\cite{bib:mixing},
can be used to determine the ratio of
CKM matrix elements $\Vub/\Vtd$
with roughly 4\% theoretical uncertainties~\cite{bib:pdg2004,bib:Kronfeld},
dominated by the uncertainties on 
square root of the bag parameter $\sqrt{B_B}$~\cite{bib:Kronfeld}.

No evidence of the $\btaunu$ decay has been reported to date.
The most stringent published experimental limit is
$\BRbtaunu < 4.2 \times 10^{-4}$ at the $90\%$ confidence level 
(C.L.)~\cite{bib:babar_prl_btn}.
Physics beyond the SM, such as supersymmetry or 
two-Higgs-doublet models, 
could enhance $\BR(\btaunu)$ up to the current
experimental limits~\cite{bib:higgs}.


The data used in this analysis were collected with the
\babar\ detector~\cite{bib:babar} at the
PEP-II asymmetric-energy $e^{+}e^{-}$ storage ring.
The results are based on a data sample
of $\bcount$ $\B\Bbar$ events,
in an integrated luminosity
of $\onlumi$ collected at the $\Y4S$ resonance.
An additional sample of $\offlumi$ was collected at 
a center-of-mass (CM) energy approximately 40~\mev below 
the $\Y4S$ resonance. We used the latter 
sample to study continuum events, $\eeqq~(q =u$, $d$, $s$, $c)$
and $\eetautau$. 
Charged-particle tracking and $\dedx$ measurements for particle 
identification (PID) are
provided by a five-layer double-sided silicon vertex tracker
and a 40-layer drift chamber operated in
the 1.5~T magnetic field of a superconducting solenoid.
A detector of internally reflected Cherenkov light (DIRC) is
used to identify charged kaons and pions.
The energies of neutral particles are measured by an electromagnetic
calorimeter (EMC) consisting of 6580 CsI(Tl) crystals.
The magnetic flux return of the solenoid is instrumented with
resistive plate chambers in order to provide muon identification.
A full detector Monte Carlo (MC) simulation based on
{\tt EvtGen}~\cite{bib:evtgen} and 
{\tt GEANT4}~\cite{bib:geant} is used to evaluate
signal efficiencies and to identify and study background sources.
Beam-related background and detector noise samples are obtained from
random triggers at regular intervals. These samples are overlaid on the
simulated events with appropriate luminosity weighting to model these
time-varying background conditions.


Due to the presence of at least two neutrinos in the final state, 
the \btaunu\ decay lacks the kinematic constraints that are 
usually exploited in $B$ decay searches in order to reject both 
continuum and $B \Bbar$ backgrounds.
The strategy adopted to search for this decay 
is to reconstruct the $\Bub$ meson
from an $\Y4S \to \Bu \Bub$ 
event in a semileptonic final state, denoted by $\tagBlep$.
All remaining charged and neutral particles in that event,
referred to as the ``signal-side'' particles throughout this paper,
are then examined under the assumption that they are attributable to 
the decay of the accompanying $\Bu$ (``signal~\B''). 


The $\tagBlep$ is reconstructed in the decay modes 
$\tagBlep \to \dszlnu$
($\ell$ = $e$ or $\mu$).
The $D^{*0}$ is reconstructed in the modes
$D^{0}\piz$ and $\Dz \gamma$. 
The $\Dz$ is reconstructed in four decay modes:
$K^{-}\pi^{+}$, $K^{-}\pi^{+}\pi^{-}\pi^{+}$, $K^{-}\pi^{+}\pi^{0}$, and
$\KS \pi^{+}\pi^{-}$. 
All kinematic variables are calculated in the CM-frame of the $\Y4S$
unless otherwise noted.

Photon candidates are obtained from EMC clusters with laboratory-frame 
energy $\eLabGam$ greater than $30$~\mev and no associated charged track.
Photon pairs with invariant mass between 
115 and 150~\mevcc are taken as $\piz$ candidates.

The $\Dz$ candidates are reconstructed by selecting combinations of
identified pions and kaons with invariant mass within
40 \mevcc of the nominal $\Dz$ mass~\cite{bib:pdg2004},
except for the $K^{-}\pi^{+}\pi^{0}$ mode, where
this window is 70 \mevcc. 
Each $\Dz$ candidate is combined with a soft $\piz$ or $\gamma$ candidate
to form a $D^{*0}$. The \piz\ and $\gamma$ 
candidates are required to have momentum less than 450 \mevc. 
Further, the $\gamma$ candidate must have $\eLabGam>100$ \mev.
The invariant mass difference $\Delta M$
between the $D^{*0}$ and $D^{0}$ is
required to be within the range 135--150 \mevcc\ for the $D^{0} \piz$ mode,
and 130--155 \mevcc\ for the $D^{0} \gamma$ mode.

The $\tagBlep \to \dszlnu$ candidates are identified
by combining a $D^{*0}$ candidate of
momentum $p_{D^{*0}}>0.5$~\gevc 
with a lepton candidate of momentum $p_{\ell}>1.0$~\gevc.
The lepton candidate must be identified as either an electron or a muon. 
The invariant mass $m_{D^{*0}\ell}$ of the $D^{*0}\ell$ candidate 
is required to be greater than $3.0$~\gevcc.
Under the assumption that a massless neutrino is the only missing particle, 
the cosine of the angle between the directions of the 
$\tagBlep$ and the lepton--$D^{*0}$ combination is
\begin{equation}
{\cosBDsl} \equiv \frac{2\,\Ebeam \cdot E_{{D^{*0}\ell}} -m^2_{{\B}} - m^2_{{D^{*0}\ell}}}
{2\, |{\bf{p}}_{\it{D^{*0}\ell}}|  \cdot \sqrt{E^2_{\rm beam}-m^2_{\B}}   } \mbox{\hspace{0.2cm} ,}
\end{equation}
where $\Ebeam$ is the expected $\Bub$ meson energy.
The energy and momentum of the $\dsz \ell$ candidate are
$E_{{D^{*0}\ell}}$ and ${\bf{p}}_{\it{D^{*0}\ell}}$, respectively.
Correctly reconstructed candidates populate the 
range $[-1$,~$1]$, whereas combinatorial backgrounds
can take unphysical values well outside this range.
We retain $\tagBlep$ candidates in the wider interval
$| \cosBDsl | < 1.1$, allowing for
the effects of detector energy and momentum resolutions.
If more than one $D^{*0} \ell$ candidate is
reconstructed in an event, the best candidate is
selected using a likelihood based on the simulated
$\Dz$ mass and $\Delta M$ distributions.
We further require that the sum of the charges of all the particles 
in the event (``net charge'') must be equal to zero.

The $\tagBlep$ reconstruction efficiency for events containing 
a $\btaunu$ decay is determined from signal simulation after
verifying that the simulated 
\BB, \uubar, \ddbar, \ssbar, \ccbar, and \tautau 
events are consistent with data.
This procedure compensates for differences in the $\tagBlep$
reconstruction efficiency in the low-multiplicity environment 
of $\btaunu$ events compared with the generic $\Bu\Bub$ environment.
The simulated efficiency is further cross-checked
by comparing the yield of events in which a 
$B^{+} \rightarrow \overline{D}^{*0} \ell^{+} \nu_{\ell}$
decay has been reconstructed in addition to a $\tagBlep$
(``double semileptonic decays'').
In the signal simulation the $\tagBlep$ reconstruction efficiency
is $\varepsilon_{\rm{sl}} = (1.75 \pm 0.07 (\rm{stat.}) \pm 0.05 (\rm{syst.})) \times 10^{-3}$.
The $\dszlnu$, $\dsz$, and $\Dz$ branching 
fractions are factored in $\varepsilon_{\rm{sl}}$.


Events that contain a $\tagBlep$ are examined for
evidence of a  $\btaunu$ decay.  
Charged tracks and EMC clusters not already utilized for the 
$\tagBlep$ reconstruction are assumed to 
originate from
the signal candidate $\Bu$ decay.
We identify the $\tau$ lepton 
in six mutually exclusive channels:
$\enunu$, $\mununu$, $\pinu$,
$\pipiznu$, $\threepinu$, 
and ``misidentified lepton''.
The misidentified-lepton channel selects signal events
from the $\enunu$ or $\mununu$ signal decays in which the 
momentum of the $\ep$ or $\mup$ from the signal $\taup$
is too low to pass the lepton identification criteria.
The identified $\tau^{+}$ modes all together correspond to 
approximately 81\% of all $\tau^{+}$ decays~\cite{bib:pdg2004}.

Signal candidates are searched in events that are required to
possess exactly one signal-side charged track, 
except for $\threepinu$ candidate events, 
which must have three signal-side charged tracks. 
The signal track from the $\enunu$ ($\mununu$) channel
is required to be identified as an electron (a muon), 
and not to satisfy either muon (electron) or kaon PID criteria. 
In the $\pinu$,
$\pipiznu$, $\threepinu$, and misidentified-lepton channels 
the signal track(s) must not satisfy electron, muon, or kaon PID.
In addition, each signal track from the $\threepinu$ channel has to 
be identified as a pion.
For the $\pipiznu$ channel the signal track is combined
with a signal-side $\piz$ candidate, reconstructed
from a signal-side photon pair ($\eLabGam$ $>$ 50 \mev for each photon)
with invariant mass between 
100 and 160 \mevcc.
If several signal-side $\piz$ candidates are reconstructed in an event, 
the candidate with $\gamma \gamma$ invariant mass
closest to the nominal $\piz$ mass~\cite{bib:pdg2004} is chosen.
We require that the events in the $\pinu$ and 
misidentified-lepton channels contain no signal-side $\piz$ candidates.
Events in the $\pinu$ and misidentified-lepton channels
are distinguished by requiring the momentum of the 
signal track to be 
greater than 1.2 \gevc in the former, and less than 1.2 \gevc in
the latter.

Further requirements are made on the (total) momentum of the 
signal track(s) for some channels:
$p_{e^{+}} < 1.4$ \gevc for $\enunu$,
and $p_{\pi^{+} \pi^{-} \pi^{+}} > 1.0$ \gevc for $\threepinu$.
We apply constraints on the missing mass $M_{\rm{miss}}$
of the event, which 
is determined by subtracting the total four-momentum
of reconstructed tracks and neutrals from
that for the \FourS\ system.
This quantity tends to be larger for events with more neutrinos.
Signal events must satisfy
$M_{\rm{miss}} > 4 $ \gevcc for $\enunu$ and $\mununu$, 
$M_{\rm{miss}} > 3 $ \gevcc for $\pinu$, $\pipiznu$ and
misidentified-lepton, and 
$M_{\rm{miss}} > 2 $ \gevcc for $\threepinu$.

Additional kinematic constraints are applied
on the $\pipiznu$ ($\threepinu$) channel,
which proceeds mainly via intermediate $\rho^{+}$ 
($a_{1}^{+}$ and $\rho^{0}$) resonance(s).
In the $\pipiznu$ channel the invariant mass of the 
$\pip \piz$ must be between 
0.55 
and 1.0 \gevcc.
For the $\threepinu$ channel
the invariant mass of the three-pion system is
required to be within the range 1.0--1.6 \gevcc.
The $\pi^{+} \pi^{-}$ combination of the three-pion system,
with invariant mass closest to the nominal $\rho^{0}$ mass~\cite{bib:pdg2004},
is required to have momentum greater then 0.5 \gevc and 
invariant mass between 0.55 
and 1.0 \gevcc.
We further require that the cosine of the angle between 
the directions of the $\tau^{+}$
and the $\pi^{+} \piz$ ($\pi^{+} \pi^{-} \pi^{+}$),
\begin{equation}
\cos\theta_{\tau,\rm{had}} \equiv \frac{2 E_{\tau} \cdot E_{\rm{had}} - m_{\tau}^{2} - m_{\rm{had}}^{2}}{2|\bf{p}_{\tau}| \cdot |\bf{p}_{\rm{had}}|},
\end{equation}
is within 
$[-1.1$,~$1.1]$.
Here 
$E_{\rm{had}}$, $\bf{p}_{\rm{had}}$ and $m_{\rm{had}}$ are the
energy, momentum and invariant mass, respectively, of the 
$\pi^{+} \piz$ ($\pi^{+} \pi^{-} \pi^{+}$).
The energy  $E_{\tau}$ and momentum $\bf{p}_{\tau}$ of the $\tau^+$
from $\btaunu$ decay are calculated under the assumption that the 
$\Bu$ is at rest in the CM-frame.

Continuum background events 
contribute to the $\pinu$, misidentified-lepton,
$\pipiznu$, and $\threepinu$ channels.
To suppress this background we
combine five variables in a linear Fisher discriminant~\cite{bib:Fisher}:
$p_{D^{*0}}$, $p_{\ell}$, $\cosBDsl$,
the cosine of the angle between the thrust axis of 
the decay products of
$\tagBlep$ and 
the thrust axis of the rest of the event, and 
the ratio of the second and zeroth Fox-Wolfram moments
using all the particles in the event~\cite{bib:foxwolfram}.
The requirement placed on the output of the Fisher discriminant
selects about 93\% of signal events and rejects about 37\%
of continuum background events.
After this requirement the continuum background
in each channel is less than 40\% of the total background.

\begin{figure}[!htb]
\includegraphics[width=\linewidth]{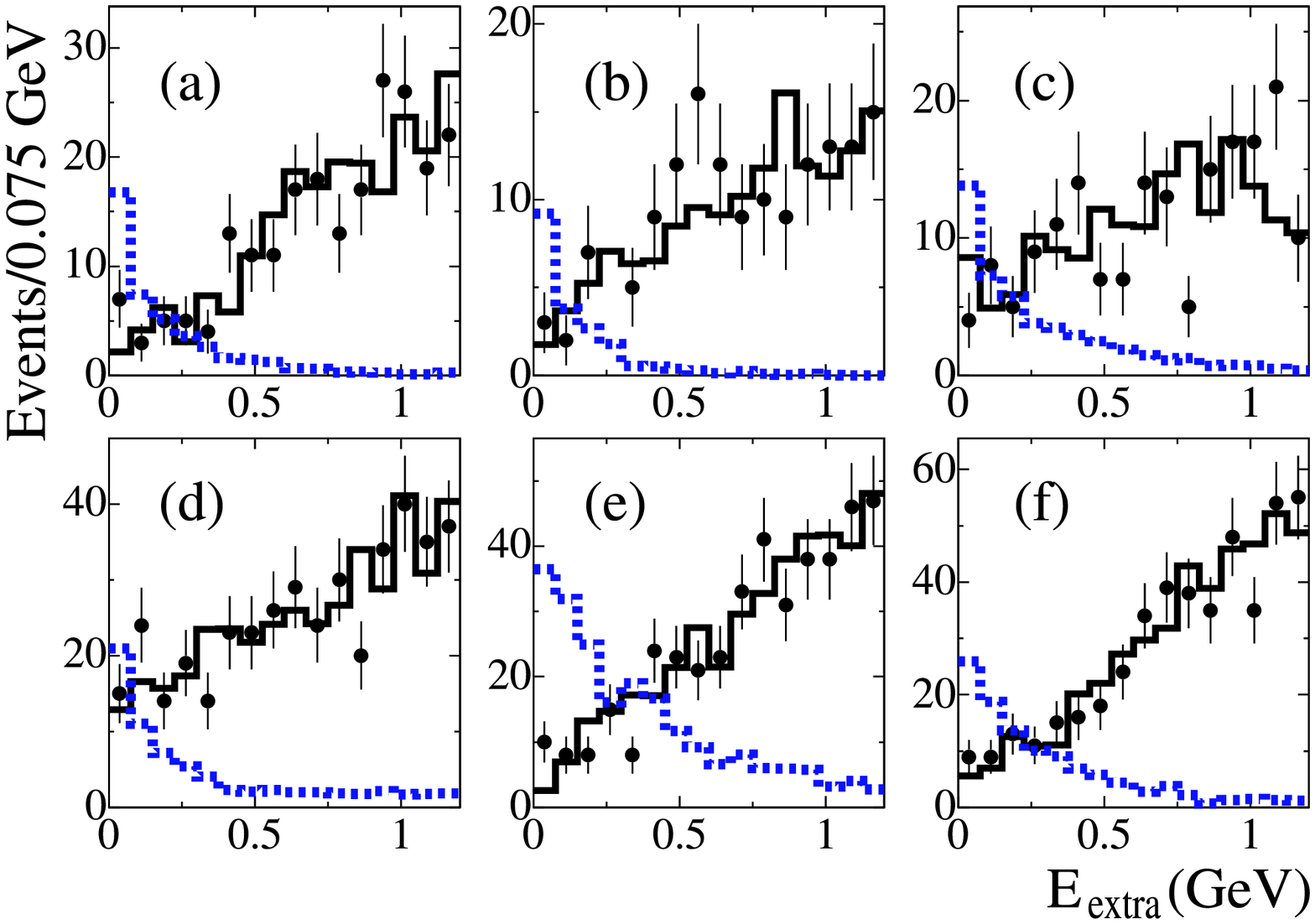}
\caption{\label{fig:eextra}%
The distribution of \eextra\
after applying all other selection criteria, plotted for 
(a) $\enunu$, (b) $\mununu$, (c) $\pinu$, 
(d) misidentified lepton, (e) $\pipiznu$, and (f) $\threepinu$
channels.
The data and background MC samples are represented by the
points with error bars and solid histograms, respectively.
The dotted lines indicate the $\btaunu$ signal distribution from MC. 
The signal MC events for the $\enunu$, $\mununu$, $\pinu$, and 
misidentified-lepton ($\pipiznu$ and $\threepinu$) 
channels are normalized assuming 
a branching fraction of $10^{-3}$ ($10^{-2}$)
for \btaunu\ decay.}
\end{figure}

The sum of the laboratory-frame energies of the neutral EMC
clusters with $\eLabGam$ $>$ 30 \mev,
which are not associated with either the 
$\tagBlep$ or the $\piz$ candidate from $\pipiznu$ 
channel, is denoted by $\eextra$ (Fig.~\ref{fig:eextra}).
For signal events the neutral clusters contributing to $\eextra$
come only from hadronic shower fragments, bremsstrahlung,
and beam-related background. 
This variable peaks near
zero for signal while for background,
which contains additional sources of neutral clusters, it 
takes on larger values.
Signal events are required to have $\eextra$ 
less than 
250 \mev for $\enunu$,
150 \mev for $\mununu$,
300 \mev for $\pinu$,
170 \mev for misidentified lepton,
250 \mev for $\pipiznu$, and 
200 \mev for $\threepinu$,
which are selected based on MC study to
provide the tightest branching fraction upper limit.
The $\eextra$ selection region
defines the ``signal region'' for each channel. 
The $350 < \eextra < 1000$ \mev region is defined as 
the ``side band'' for all the channels.

The efficiencies $\varepsilon_i$ for each $\tau$
selection channel $i$ 
are determined using simulated events.
Cross-feeds among the $\tau$-decay channels are taken into account.
The systematic uncertainties in the selection efficiency arise 
from tracking efficiency (1.4\% per track), 
particle identification (0.2\%--2.0\%),
$\eextra$ simulation (3.0\%--8.0\%), 
\piz\ reconstruction (3.3\%), and 
data and MC differences in the output of the Fisher discriminant (1.0\%).
Systematic uncertainties due to the $\eextra$ simulation are determined by
evaluating the effect of varying the MC $\eextra$ distribution within 
a range representing the observed level of agreement with data 
in samples containing $\tagBlep$ and up to seven additional tracks.
For further cross-check the $\eextra$ distributions 
of the data and MC events for the double semileptonic decays
are compared.
The signal selection efficiencies for the six selection channels 
are listed in Table~\ref{tab:SigSideSelEff}. The total $\btaunu$
selection efficiency is roughly 31\%.

The remaining background consists primarily of $\Bu\Bub$ events 
with correctly reconstructed $\tagBlep$. For these events
the signal side contains
$K_{L}^{0}$'(s), neutrino(s), or particles that pass outside the
detector acceptance. For each channel 
we estimate the background $b_i$ in the signal region
using events in the data side band 
and the simulated $E_{\rm{extra}}$ distribution:
\begin{equation}
b_{i} =  N_{\rm{SideB}}^{\rm{data}} \times ( N_{\rm{SigR}}^{\rm{MC}} / N_{\rm{SideB}}^{\rm{MC}} ). 
\end{equation}
Here $N_{\rm{SideB}}^{\rm{data}}$ is the 
number of data events in the side band, and 
$N_{\rm{SigR}}^{\rm{MC}}$ and $N_{\rm{SideB}}^{\rm{MC}}$
are the numbers of MC background events in the signal region and side band, 
respectively. Background estimation is cross-checked using
data and MC events that satisfy the full signal selection, 
with the exception of having two signal-side tracks, or 
non-zero net charge,
or the $\Delta M$ of the $\dsz$ outside the selection region.
The uncertainties in the background estimations are predominantly 
statistical; smaller 
systematic uncertainties arise from the simulation of the
$\eextra$ shape in the background MC.

We determine the \btaunu\ branching fraction from the number of signal
candidates $s_i$ expected for each $\tau$ selection channel, 
where $s_i \equiv
N_{\B^{\pm}} \, \varepsilon_{\rm{sl}} \, \varepsilon_i \, \BRbtaunu$.
$N_{\B^{\pm}} = \bcount$ is the estimated number of $\B^{\pm}$ mesons
in the data sample.
The results for each channel are combined using the estimator
$Q \equiv {\cal L}(s+b)/{\cal L}(b)$~\cite{bib:cls,bib:lista}, where
\begin{equation}
  {\cal L}(s+b) \equiv
  \prod_{i=1}^{6}\int_{-\infty}^{+\infty} {\mathrm d}b_i^\prime \, \,
  \frac{e^{-\frac{(b_i^\prime - b_i)^2}{2\sigma_i^2}} }{ \sqrt{2\pi \sigma_i^2} } \,
  \frac{ e^{-(s_i+b_i^\prime)}(s_i+b_i^\prime)^{n_i}}{ n_i!}
  \label{eq:lb}
\end{equation}
is the likelihood function for signal-plus-background hypotheses,
$n_i$ is the observed number of data events in each $\tau$ selection channel,  
and $\sigma_i$ is the uncertainty in the background 
estimate $b_i$ (Table~\ref{tab:SigSideSelEff}).
The likelihood function for background-only hypotheses ${\cal L}(b)$
can be obtained from Eq.~\ref{eq:lb} by setting $s_i$ to zero.

\begin{table}[t]
\caption{\label{tab:SigSideSelEff}
Efficiency ($\varepsilon_i$) with statistical and systematic errors, 
expected background ($b_i$), and observed data candidates ($n_i$) for 
each reconstructed $\tau$ selection channels. 
The cross-feeds among the $\tau$ decay modes are taken into account.
The $\varepsilon_i$ values include the branching fractions of 
the $\tau$ decay modes.}
\begin{center}
\begin{tabular}{l c c c} \hline \hline
selection      & $\varepsilon_i (\%)$      &  $b_i$             & $n_i$ \\ \hline
$\enunu$       & $\:\:7.5 \pm 0.4 \pm 0.2$ &  $\:13.4  \pm 2.4$ & $\:17$\\
$\mununu$      & $\:\:2.9 \pm 0.2 \pm 0.1$ &  $\:\:6.2 \pm 1.7$ & $\:\:5$\\
$\pinu$        & $\:\:8.0 \pm 0.4 \pm 0.3$ &  $\:27.7  \pm 5.0$ & $\:26$\\
$\pipiznu$     & $\:\:2.5 \pm 0.2 \pm 0.1$ &  $\:28.6  \pm 4.3$ & $\:31$\\
$\threepinu$   & $\:\:1.4 \pm 0.2 \pm 0.1$ &  $\:21.6  \pm 3.0$ & $\:26$\\
misidentified lepton & $\:\:9.0 \pm 0.4 \pm 0.4$ &  $\:33.4  \pm 5.1$ & $\:45$\\

 \hline \hline
\end{tabular}
\end{center}
\end{table}

The measured branching fraction, which is the value that maximizes
the likelihood ratio estimator, is 
$(1.3^{+1.2}_{-1.1}) \times 10^{-4}$.
This value is compatible with a zero branching fraction.
The $n_i$ and $b_i$ values
(Table~\ref{tab:SigSideSelEff})
do not indicate any significant excess of observed events.
Therefore, we set an upper limit on
the branching fraction~\cite{bib:lista} of
$\BRbtaunu < 2.8 \times 10^{-4}$~($90\%$ C.L.). 
The expected branching fraction upper limit
for background only hypothesis is 
$\BRbtaunu < 1.8 \times 10^{-4}$~($90\%$ C.L.).


The \babar\ Collaboration has previously performed 
a search for the $\btaunu$ decay
based on a sample of $88.9 \times 10^{6}$ $\BB$ pairs,
where the \Bub\ meson accompanying the signal \Bu\
is reconstructed in a variety of hadronic or semileptonic 
modes~\cite{bib:babar_prl_btn}. 
The hadronic $\Bub$ selection is mutually exclusive with the
current $\tagBlep$ selection. Therefore
the two samples are statistically independent and
may be combined. The hadronic reconstruction analysis 
obtained a limit
$\mathcal{B}(\btaunu) < 4.2 \times 10^{-4}$ at the $90\%$ C.L.
To combine the results from the previous hadronic and 
current semileptonic samples,
we create a combined estimator from the product
of the semileptonic ($Q_{{\rm sl}}$) and 
hadronic ($Q_{{\rm had}}$) likelihood ratio
estimators, $Q \equiv Q_{{\rm sl}} \times Q_{{\rm had}}$.
The measured branching fraction from the combined sample
is 
$(1.3^{+1.0}_{-0.9}) \times 10^{-4}$.
This value is compatible with a zero branching fraction, and
we set a combined upper limit,
\begin{equation}\label{eqn:thelimit}
\BRbtaunu < 2.6 \times 10^{-4} \, \textrm{($90\%$ C.L.).} 
\end{equation}
These results represent the most stringent
limits on $\btaunu$ reported to date.

We are grateful for the excellent luminosity and machine conditions
provided by our \pep2\ colleagues, 
and for the substantial dedicated effort from
the computing organizations that support \babar.
The collaborating institutions wish to thank 
SLAC for its support and kind hospitality. 
This work is supported by
DOE
and NSF (USA),
NSERC (Canada),
IHEP (China),
CEA and
CNRS-IN2P3
(France),
BMBF and DFG
(Germany),
INFN (Italy),
FOM (The Netherlands),
NFR (Norway),
MIST (Russia), and
PPARC (United Kingdom). 
Individuals have received support from CONACyT (Mexico), A.~P.~Sloan Foundation, 
Research Corporation,
and Alexander von Humboldt Foundation.


\begin{thebibliography}{99}
\bibitem{bib:cc}
Charge-conjugate modes are included implicitly throughout this paper.
\end{thebibliography}
\end{document}